\documentclass[%
 aip,
 rsi,
 amsmath,amssymb,
 reprint,%
]{revtex4-1}

\usepackage{graphicx}
\usepackage{dcolumn}
\usepackage{bm}

\usepackage[utf8]{inputenc}
\usepackage[T1]{fontenc}
\usepackage{mathptmx}

\begin{document}

\preprint{AIP/123-QED}

\title{Free-space confocal magneto-optical spectroscopies at milliKelvin temperatures}

\author{B.J. Lawrie}
\email{lawriebj@ornl.gov; This manuscript has been authored by UT-Battelle, LLC, under contract DE-AC05-00OR22725 with the US Department of Energy (DOE). The US government retains and the publisher, by accepting the article for publication, acknowledges that the US government retains a nonexclusive, paid-up, irrevocable, worldwide license to publish or reproduce the published form of this manuscript, or allow others to do so, for US government purposes. DOE will provide public access to these results of federally sponsored research in accordance with the DOE Public Access Plan (http://energy.gov/downloads/doe-public-access-plan).}
\affiliation{ 
Materials Science and Technology Division, Oak Ridge National Laboratory, 1 Bethel Valley Rd, Oak Ridge, TN 37831 USA}%
\affiliation{Quantum Science Center, Oak Ridge, TN 37831 USA}
\author{M. Feldman}
\affiliation{ 
Materials Science and Technology Division, Oak Ridge National Laboratory, 1 Bethel Valley Rd, Oak Ridge, TN 37831 USA}%
\affiliation{Department of Physics and Astronomy, Vanderbilt University, Nashville TN 37235 USA}
\author{C.E. Marvinney}
\affiliation{ 
Materials Science and Technology Division, Oak Ridge National Laboratory, 1 Bethel Valley Rd, Oak Ridge, TN 37831 USA}%
\author{Y.Y. Pai}
\affiliation{ 
Materials Science and Technology Division, Oak Ridge National Laboratory, 1 Bethel Valley Rd, Oak Ridge, TN 37831 USA}%
\affiliation{Quantum Science Center, Oak Ridge, TN 37831 USA}

\date{\today}

\begin{abstract}
We describe the operation of a free-space confocal optical microscope operated in a dilution refrigerator.  The microscope is installed on a cold insertable probe to enable fast sample exchange while the refrigerator is held at low temperatures. A vector magnet provides a 6 T field normal to the sample and 1 T fields at arbitrary angles. A variety of optical microscopies and spectroscopies, including photoluminescence, Raman, magneto-optical Kerr effect, and spin relaxometry measurements are described, and some of the challenges associated with performing these measurements at milliKelvin temperatures are explored.
\end{abstract}

\maketitle

\section{Introduction}
A growing number of quantum technologies rely on low temperature operation at 10 mK-1 K. For instance, the recent demonstration of quantum supremacy with a quantum computer based on superconducting qubits\cite{neill2018blueprint,arute2019quantum} relied on decades of development of superconducting devices at mK temperatures. Similarly, superconducting nanowire single photon detectors are now widely used for applications ranging from quantum sensing\cite{hochberg2019detecting,ahmed2018quantum} to quantum networking\cite{hadfield2006quantum,takemoto2015quantum,zhang2018experimental,sun2016quantum,graffitti2020measurement} because they offer a combination of high-speed, high quantum efficiency, and low dark-count-rate single photon detection at spectral bands beyond those covered by conventional semiconducting detectors and temperatures of order 1 K\cite{natarajan2012superconducting,eisaman2011invited}. Transition edge sensors operated at mK temperatures offer photon number resolution that is critical to fundamental tests of quantum information \cite{vaidya2020broadband,nehra2019state,smith2012conclusive}. Some spin states that are drawing interest as possible qubits like the silicon vacancy center in diamond must operate at mK temperatures to mitigate against phonon-induced spin decoherence processes~\cite{Sukachev2017}. Further, topological devices that may support Majorana fermions at mK temperatures are drawing interest as potential platforms for quantum computation~\cite{mourik2012signatures,das2012zero}.

Scanning laser microscopies have been used extensively over the past two decades to characterize phase-slip behavior, hot spot dynamics and edge superconductivity in superconducting devices~\cite{werner2013edge,wang2009hot,sivakov2003josephson}. Over the past several years, there has been a strong interest in developing single photon imaging capabilities with large area and multi-pixel SNSPDs~\cite{marvinney2020waveform,wollman2019kilopixel,dauler2007multi,zhao2017single}.  And there are now many examples of magneto-Raman, and photoluminescence microscopies at cryogenic temperatures~\cite{li2018probing,jiang2019spin,mccreary2020distinct}. However, most of these experiments have been performed at temperatures above 1 K. 

Optical microscopies capable of manipulating and probing quantum materials and quantum devices at milliKelvin temperatures are  critical to the development of the next generation of quantum devices, but such microscopies have historically been very limited. The laser power required for typical experiments is often of order the cooling power of commercially available dilution refrigerators, limiting the accessible base temperature for many measurements. Injecting light into a dilution refrigerator via free space optics introduces challenges associated with blackbody radiation and long optical path lengths\cite{Sukachev2017}. Fiber-coupled microscopes generally offer limited imaging capabilities as well as challenges with blackbody radiation~\cite{sladkov2011polarization,macdonald2015optical}. Closed cycle dilution refrigerators tend to introduce substantial levels of vibration that can limit spatial resolution. Further, measurements that rely on large magnetic fields can prove challenging to interpret because of substantial Faraday rotation acquired within the microscope optics. 

Here, we describe the design and operation of a free-space optical microscope in a closed cycle dilution refrigerator with an eye toward addressing the above technical challenges in order to characterize and control quantum materials and quantum devices at the mesoscale. In particular, we describe the accessible base temperature for different classes of experiments (including photoluminescence, Raman, relaxometry, and magneto optical Kerr effect (MOKE) microscopies) that require different laser powers, we describe an approach to sample positioning that combines 3-axis stick-slip positioners with an 8f imaging system and a 2-axis galvo scanner, we describe Faraday rotation compensation for magneto-optical spectroscopies, and we discuss the integration of superconducting single photon detectors in the dilution refrigerator.

\section{Dilution Refrigerator}

The microscope is based on a Nanomagnetics Instruments confocal microscope installed in a Leiden Cryogenics CF-CS81-1000M closed cycle dilution refrigerator. The dilution refrigerator includes an American Magnetics 6/1/1 T superconducting vector magnet with a 90 mm bore. The microscope head is installed on a 81 mm diameter cold insertable probe that allows for rapid sample transfer within 1 day while keeping the dilution refrigerator at ~4 K (at the expense of reduced cooling power and increased base temperature on the probe compared with the dilution refrigerator mixing chamber).  With the cold insertable probe removed and a radiation shield installed in its place, the measured base temperature of the mixing chamber is 10.5 mK, and the base temperature with 1 mW of applied heat is 104.6 mK.  With the cold insertable probe installed, the base temperature of the mixing chamber climbs to 17.8 mK and the base temperature of the probe is 20.4 mK. Under normal conditions the cooling power for the probe is 128 $\mu$W at 100 mK. Much of the design of the optical microscope is centered on this understanding of the available cooling power.

The base temperature of the microscope mounted on the cold insertable probe is highly dependent on the thermal conductivity between each thermal stage of the refrigerator and the probe. Under normal operation, the cold insertable probe is clamped to each thermal stage of the refrigerator using a pneumatic actuator.  The cooling power of the probe varies with time as contamination builds up between the probe clamps and the refrigerator plates.  Infrequently warming up the refrigerator and pumping the inner vacuum chamber resolves this issue and re-optimizes the cooling power on the probe.


\section{Vibration control}
Closed cycle refrigerators based on pulse tube cryocooling are now ubiquitous in part because of the cost and experimental downtime associated with liquid helium-based systems~\cite{de2011basic}. However, pulse-tube cryocoolers deliver substantial low frequency impulses directly to the cryostat, and they can additionally impart helium gas mediated acoustic frequency vibrations~\cite{olivieri2017vibrations,caparrelli2006vibration}.  As a result, most cryogenic scanning probe and scanning tunneling microscopies rely on liquid cryogens.  However, some effort has been invested over the past decade in the development of vibration control for pulse-tube-cryocooler-based closed-cycle mk scanning probe microscopies~\cite{den2014atomic,pelliccione2013design}. Controlling these vibration levels is critical even to mesoscale microscopies and to quantum device operation because the micron-scale amplitude of vibrations in typical closed cycle dilution refrigerators can induce electrical noise through the triboelectric effect that can lead to increased decoherence~\cite{kalra2016vibration}.

Recent demonstrations of closed-cycle milliKelvin scanning probe microscopies have relied on a combination of strategies to reach sub-Angstrom noise levels, including: (1) mechanical decoupling of the pulse-tube cooler from the dilution refrigerator, (2) massive mixing chambers and microscope stages, and (3) passive damping with mechanical springs~\cite{den2014atomic,pelliccione2013design}. These choices have enabled atomic resolution at the expense of thermal anchoring and fast sample exchange.  Cold-insertable probes are generally incompatible with flexible links within the dilution refrigerator, and reduced mechanical stiffness typically results in reduced thermal anchoring. MilliKelvin optical microscopies do not necessarily require Angstrom-level vibration tolerances, but minimizing the vibration levels in our microscope is critical to achieving diffraction limited spatial resolution, minimizing associated electronic noise, and minimizing acoustic noise in optical spectroscopies.

Here, we have compromised between these previously used approaches and the requirement for fast sample exchange while introducing active vibration isolation. In order to allow for the use of the cold-insertable probe, we use a conventional rigid framework inside the dilution refrigerator. However, inspired by other groups' reports~\cite{den2014atomic}, we reduced the mechanical coupling between the pulse tube and the room temperature, 50 K, and 3 K plates by lifting the pulse tube motor off of the room temperature stage, and suspending it with rubber straps from a floor-mounted unistrut manifold. In addition, we mount the room temperature stage on a pair of Halcyonics Vario series active vibration isolation elements, which help to decouple the dilution refrigerator from vibrations in the room, including those generated by the adjacent gas handling system.

\begin{figure}[t]
\centering
    \includegraphics[width=\columnwidth]{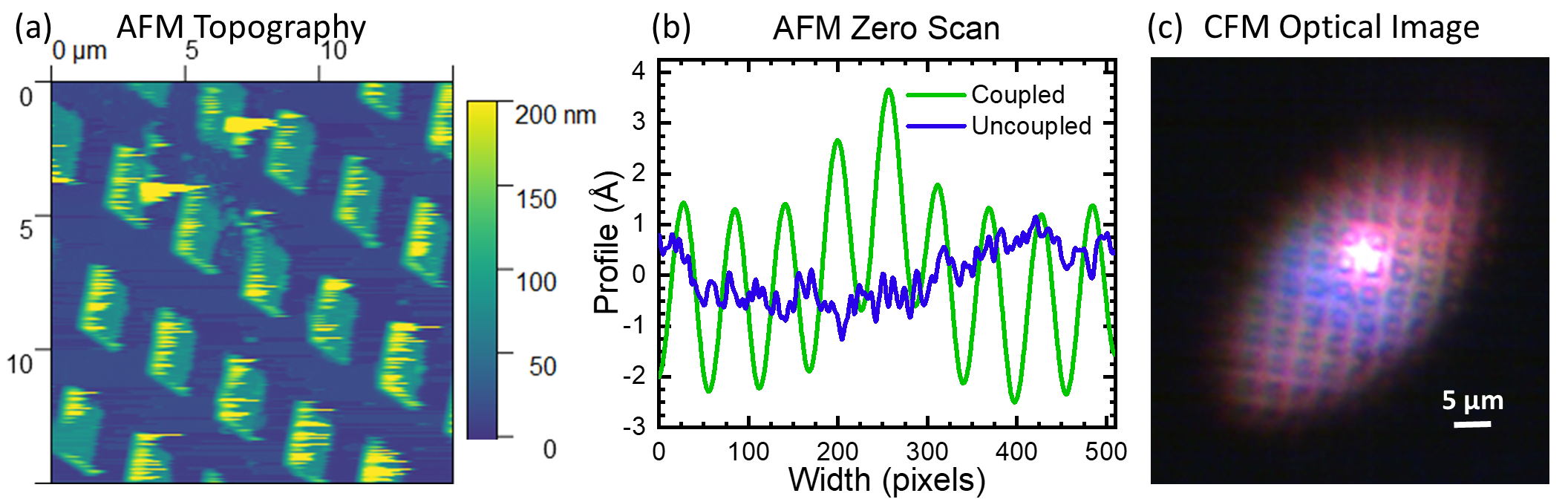}
    \caption{Representative images collected from microscopes mounted to the cold-insertable probe. a) AFM contact-mode image of a calibration sample grating with 2 micron pads collected at 500 mK.  b) AFM zero scans illustrating noise levels at the mixing chamber when the Halcyonics vibration isolation is on. (blue) the pulse tube motor is decoupled from the 300 K plate and the microscope is held at 500 mK.  (green) the pulse tube motor is directly mounted to the 300 K plate and the microscope is held at 4 K.  c) A confocal microscope image of the same grating, taken during liquid nitrogen precooling at 189 K with the pulse tube motor decoupled from the 300 K plate and active vibration isolation in place.  A red (635 nm) alignment laser is focused on the sample between the pillars.
}
    \label{fig:vibrations}
\end{figure}

In order to test the vibration levels in our microscope, we installed a Nanomagnetics Instruments atomic force microscope (AFM) on the mixing chamber and measured the vertical noise levels with and without the active vibration isolation system and with and without the pulse-tube-motor decoupling from the 300 K plate.  Figure~\ref{fig:vibrations}a illustrates a typical AFM image acquired for a CCD array with 2 micron pads. That measurement was taken at 500 mK with the active vibration isolation system in use, and the pulse tube motor decoupled from the 300 K plate.  Figure~\ref{fig:vibrations}b illustrates the measured zero-scan noise floor with the active vibration isolation system turned on.  When the pulse tube motor was decoupled from the 300 K plate, RMS noise levels of 2.40 angstroms were measured, with a linescan of the vibrations shown in the blue curve.  When the pulse tube motor was bolted to the 300 K plate, the AFM zero-scan exhibited substantial harmonic motion as well as a slightly larger RMS noise level of 3.11 angstroms, with a linescan of the vibrations shown in the green curve.  When the active vibration isolation system was turned off, we were unable to operate the AFM as a result of excess noise levels.  An optical image of the same grating taken in the confocal microscope during precooling (with both liquid nitrogen and the pulse tube compressor on) is shown in Fig.~\ref{fig:vibrations}c for reference. Notably, these measurements do not describe the relative vibrations between the optical breadboard mounted to the room-temperature stage and the sample mounted in the microscope at the mixing chamber stage.  We have not recorded any measurements of the relative vibrations between the breadboard and the mixing chamber except to qualitatively note that micron scale vibrations are present in optical microscopy images when the Halcyonics vibration isolation stages are turned off and the pulse-tube motor is mounted on the 300 K plate.  Those vibrations are no longer resolvable when the active vibration isolation is turned back on and the pulse tube motor is suspended from its unistrut frame.

\section{Development of a mK-Scanning Confocal Microscope}

\begin{figure}[h]
\centering
    \includegraphics[width=\columnwidth]{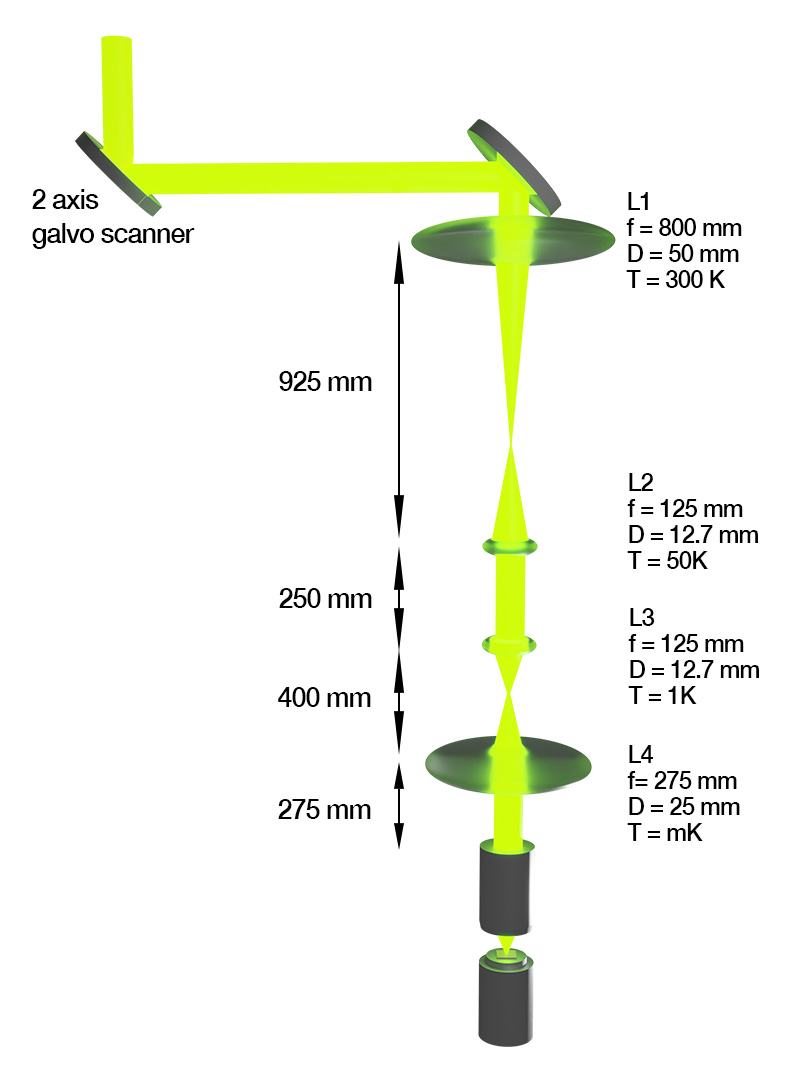}
    \caption{8F scanning imaging system with the positions of the four lenses L1-L4.}
    \label{fig:fig8Fscan}
\end{figure}

Many efforts focused on the characterization of nanophotonic systems at mK temperatures rely on fiber-optic delivery in order to minimize blackbody radiation, reduce detrimental vibration effects, and simplify the optics train.  However, free-space optical delivery is essential to many applications that require high spatial resolution imaging and manipulation and flexible control of wavelength and polarization states. As illustrated in Figure~\ref{fig:fig8Fscan}, we rely on free space delivery to the microscope through an 8f imaging system installed on the 8mm clear bore in the center of the cold insertable probe. Optical windows heatsunk to each stage of the refrigerator have been used in the past to filter out black body radiation when free-space optical access is required.  Here we found no difference in base temperature or cooling power when we exchanged those optical windows for the heatsunk lenses required for the 8f imaging system. While we do not have lenses mounted to all five stages of the refrigerator, this layout appears to block enough black body radiation while simultaneously providing improved imaging functionality. 

At the bottom of the mixing chamber, a NA=0.84 cryogenic objective is mounted on a z-axis stick-slip positioner. Below that is a three axis xyz-stick-slip positioner for sample positioning. The sample stage is heatsunk to the mixing chamber via thermal braids. However, the stick-slip positioners can heat the sample by up to a few hundred milliKelvin (for large step sizes) and introduce rf artifacts, so the stick slip positioners are generally used for coarse sample positioning but not for imaging. The 8F imaging system described above and illustrated in Figure~\ref{fig:fig8Fscan} coupled to galvanometer controlled scanning mirrors provides reasonable scanning capability with none of the detrimental effects of the stick-slip positioners. Here, four lenses L1-L4 are placed on the cold-insertable probe: one at atmosphere outside the refrigerator, one on the 50 K stage, one on the 1 K stage and one on the mixing chamber stage. Using a ray optics simulator, Zemax, we simulate a scan range of $\pm$36$\mu$m.  For a 740 nm laser, the modeled spot size increases from 696 nm to 699 nm as the scanner moves from the center of the image to the edge of the image.  Our scan range is ultimately limited by the 8 mm diameter clear aperture of our 1 meter long radiation shield on the cold insertable probe. A larger scan range could be achieved with a larger bore, but it is difficult to integrate a larger bore into the 81 mm diameter probe without negatively impacting the structural integrity of the probe and the space needed for coaxial and fiber delivery to the sample. Previous research efforts that relied on larger bore galvo scanning solutions have achieved larger scan ranges at the expense of working without a cold insertable probe~\cite{Sukachev2017}.

\begin{figure}[b]
\centering
    \includegraphics[width=\columnwidth]{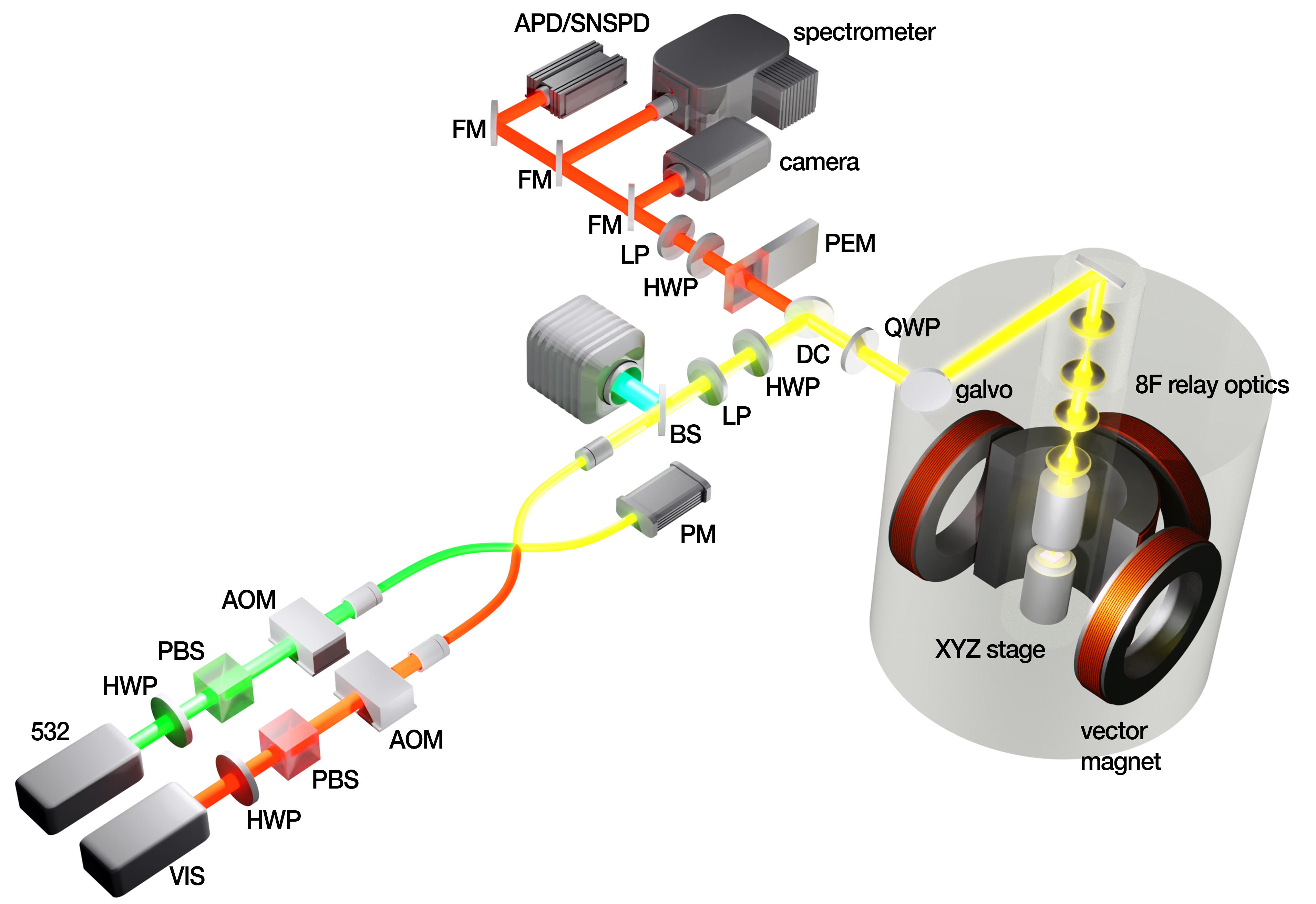}
    \caption{Schematic of optics train for mK-scanning confocal microscope. Half-wave plates (HWPs) and polarization beam splitters (PBSs) are used to control the power of a 532 nm laser and a tunable optical parametric oscillator (Spectra Physics Millenia Edge and Hubner Photonics C-Wave, respectively).  These two light sources are coupled into a pair of acousto-optic modulators (AOMs), combined in a fiber beamsplitter and delivered to the excitation optics on the refrigerator breadboard via a single mode fiber. The second arm of the fiber beamsplitter is used to monitor the laser power (with digital control of the AOMs used to select emission in one or both colors).  A LED is combined with the laser sources on a beamsplitter (BS). The optical polarization is initialized using a linear polarizer (LP) and a half-wave plate or a quarter-wave plate (QWP) as appropriate. The excitation arm is reflected off a low-pass dichroic mirror (DM) to a 2-axis galvonometer-controlled mirror which reflects the laser light onto the 8F imaging system, the 0.85 NA objective, and the sample. Light from the sample is collected by the objective and by reciprocity is passed to the DM where it is transmitted to a spectrometer, camera or single photon detectors using a sequence of electronically controlled flip mirrors (FMs). An additional linear polarizer or photoelastic modulator (PEM) is inserted into the collection arm as appropriate.}
    \label{fig:schematic}
\end{figure}

Given the 80 cm path length between the first lens on the probe and the galvo scanner, it proved essential to incorporate some of the microscope optics train onto a breadboard rigidly mounted to the top of the dilution refrigerator as shown in figure~\ref{fig:schematic}.  This breadboard is roughly 2.1 m above the lab floor, so all laser sources are delivered to the breadboard by single mode fiber, and all single photon detection and spectroscopy is performed after collection into multimode fibers with core size of 30-100 $\mu m$. The specific optical train layout varies slightly between experiments, but the basic optical train is described here.

Optical excitations (from a continuous wave 532 nm laser, an optical parametric oscillator tunable from 450-650 nm and 900-1300 nm, or a supercontinuum laser) are delivered to the breadboard via single mode fiber and reflected from a dichroic mirror (DM) that reflects the laser onto the galvo-controlled mirror and 8F imaging system. Photons scattered or emitted by the sample are then transmitted through the dichroic mirror (DM) and long pass filter (LP).  Automated flip mirrors are used to direct the collected light to a camera, single photon detectors, or a spectrometer. Polarization optics are optionally included in the excitation and collection, and are described in further detail below.

\section{Polarization control}

The ability to perform magneto optical spectroscopies - including polarization resolved Raman and photoluminescence spectroscopies, MOKE microscopies, and spin noise spectroscopies - hinges on a clear understanding of the polarization state of light within the dilution refrigerator.  All of these spectroscopies and microscopies are generally optimally performed with a microscope objective that is mounted in close proximity to the sample. Unfortunately, in high magnetic fields, most transmissive optical components induce substantial Faraday rotation of the polarization state. The microscope objective is also in a 6 T field when the sample is held at 6 T.  The other lenses in the 8f imaging system illustrated in Fig.~\ref{fig:fig8Fscan} induce reduced Faraday rotation compared with the objective, but L4 is in a roughly 30 mT field when the sample is held at 6 T.  For a 6 T field at the sample, the field at L3 is reduced to 6 mT, and the field at L2 is roughly 3 mT. The field at the room temperature window and L1 is below 0.5 mT. The Faraday rotation resulting from these optics can therefore be calibrated for linearly polarized light by characterizing the polarization state of light reflected out of the microscope by a nonmagnetic bulk substrate like a silicon chip or a gold thin film.  However, the Verdet constant for typical optical glasses is not typically constant across the near-UV to near-IR spectrum, so it is important to characterize the Faraday rotation as a function of wavelength.  Moreover, using galvo scanning for optical microscopies as described above results in changes to the optical path length that affect both the Faraday rotation within the objective and the collection efficiency.  Thus, any polarization rotation calibration needs to be performed separately for each galvo position. 

For experiments that rely on circularly polarized light, including magnetic circular dichroism microscopies, this calibration process becomes less problematic because equal Faraday rotation is observed by both in-plane polarization components, resulting in no phase shift and no change to the circular polarization state. However, dielectric mirrors are used where possible in the optics train in order to minimize optical losses.  While dielectric mirrors offer 1-2\% reduced loss per optic compared with silver mirrors across the visible spectrum, they alter the state of circularly polarized light in a way that metallic mirrors do not.  For non-normal incidence, horizontally or vertically polarized light incident on a dielectric mirror will maintain its polarization state but acquire a change in phase.  Because of this change in phase, circularly polarized light will be converted into elliptically polarized light with the ellipticity determined by the angle of incidence and the wavelength of light. Utilizing silver mirrors in the collection arm between the sample and the LP allows us to perform magneto-optical spectroscopies with circularly polarized light while still benefiting from dielectric mirrors elsewhere in the optics train.

\begin{figure}[h]
\centering
    \includegraphics[width=.9\columnwidth]{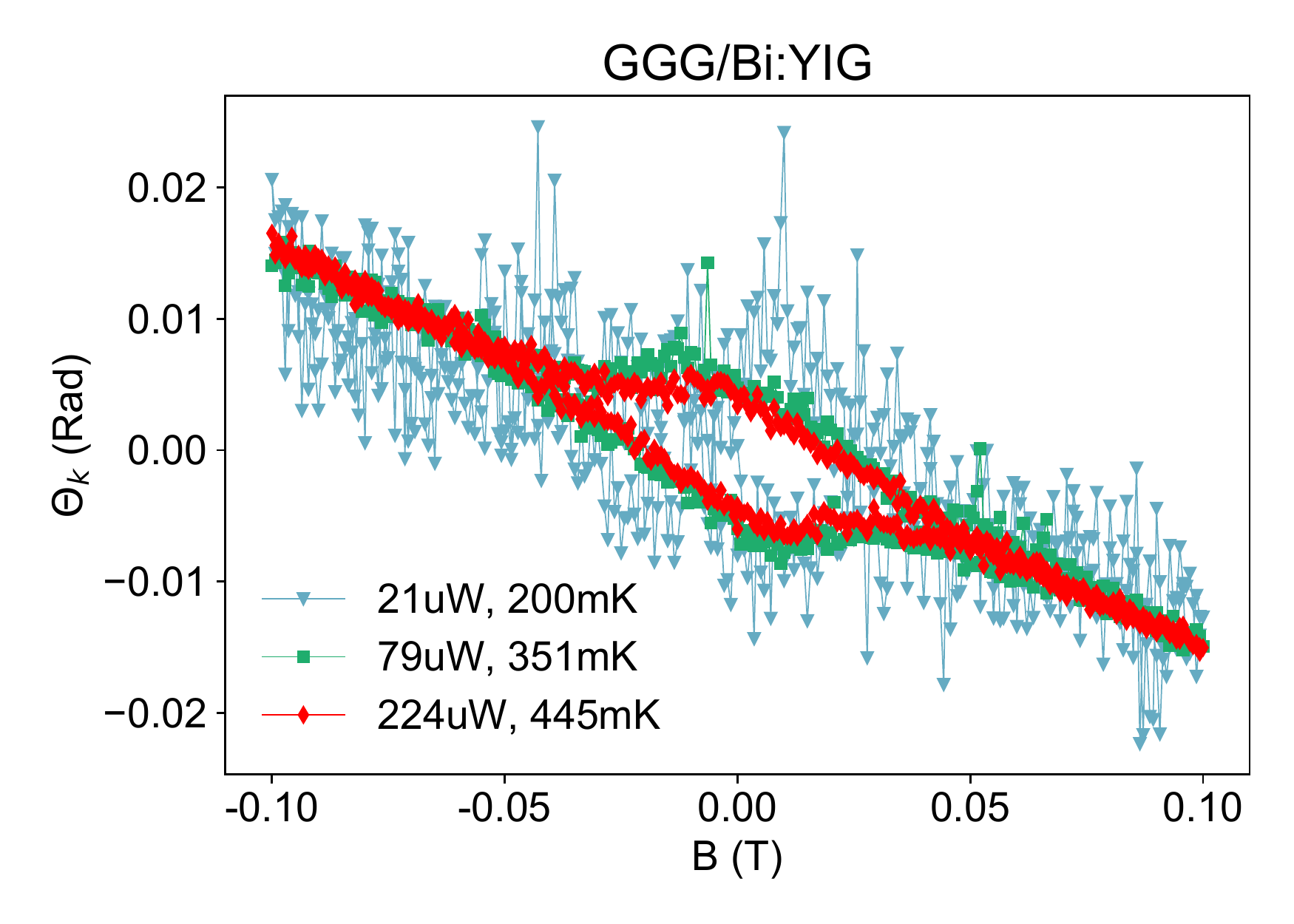}
    \caption{ MOKE characterization of a Bi:YIG thin film grown on a GGG substrate. Gold is then deposited onto the Bi:YIG as a reflective surface. The sample is mounted with Au surface facing down. Continuous wave 532 nm laser powers of 21 $\mu W$, 79 $\mu W$, and 224 $\mu W$ resulted in base temperatures of 200 mK, 351 mK, and 445 mK during this measurement. The linear background is a result of Faraday rotation in the objective and the GGG substrate.}
    \label{fig:polarization}
\end{figure}

Figure~\ref{fig:polarization} illustrates a MOKE measurement for a Bi:YIG thin film grown on a GGG substrate for a variety of laser powers. In each case, no resistive heating was applied to the sample, and the sample temperature was determined by the choice of laser power.  Notably, the heatsinking between the dilution refrigerator and the cold-insertable probe was not optimized for this measurement, resulting in reduced cooling power compared to that reported above. Optimizing the heatsinking regularly (by adjusting the mechanical clamping and by warming the dilution refrigerator and pumping out contaminants that reduce the thermal conductivity between the probe and the dilution refrigerator) help to optimize the cooling power and available base temperature for experiments, but the minimum base temperature is ultimately determined by the number of pump photons required to achieve a practical signal-to-noise ratio (SNR). This tradeoff between SNR and sample temperature is an obstacle for all mK optical spectroscopies.  Most spectroscopies can be performed at lower temperatures with no change in signal-to-noise ratio by operating at lower laser powers with longer integration times, but experiments that rely on extensive parameter sweeps quickly become infeasible as integration times increase. Notably, the hysteretic MOKE response illustrated in Fig.~\ref{fig:polarization} is a direct measurement of the sample response, but the linear slope of -0.20 rad/T is a result of Faraday rotation in the objective.  Smaller 1T fields perpendicular to the optical axis induced negligible Faraday rotation by comparison. 

\section{Single-photon detection}
A variety of cryo-optical microscopies rely on single-photon detection. An ideal single-photon detector has $100\%$ quantum efficiency, 0 dark counts per second, dead time of 0 seconds, and a timing jitter of 0 seconds. For a variety of quantum microscopies, large area multimode detectors are essential to optimizing the extrinsic quantum efficiency. Moreover, for many applications in photonic quantum information processing, photon number resolution is essential. Avalanche photon detectors (APDs), photomultiplier tubes (PMTs) and superconducting nanowire single photon detectors (SNSPDs) are widely used as single photon detectors at ultraviolet to infrared wavelengths, and transition edge sensors (TESs) are increasingly used when photon number resolution is required~\cite{eisaman2011invited}. APDs and PMTs are generally suitable for many applications at visible wavelengths.  They operate at room temperature with quantum efficiency of order $10-60\%$, dark count rates of less than 1000 per second, jitter of order hundreds of picoseconds, and maximum count rates of order $10^7$ per second.  They do not generally offer photon number resolving functionality, and at near-infrared wavelengths, their quantum efficiency drops substantially while their dark count rates climb~\cite{eisaman2011invited}. SNSPDs offer substantially better quantum efficiency, jitter, and dark count rates than APDs or PMTs at infrared wavelengths, and they maintain moderately better quantum efficiency, an order of magnitude better jitter, and 1-2 orders of magnitude better dark count rates in the visible. Moreover, the SNSPD waveform risetime can be used to provide limited spatial-resolution and photon-number resolution \cite{cahall2017multi,marvinney2020waveform}. However, SNSPDs typically rely on operating temperatures of order 1 K, and TESs operate at roughly 100 mK. As a result, SNSPDs and TESs are increasingly used in applications where APDs and PMTs are ineffective, while APDs and PMTs are still frequently used for visible and near-UV measurements where photon number resolution is not required.

\begin{figure}[h]
\centering
    \includegraphics[width=.9\columnwidth]{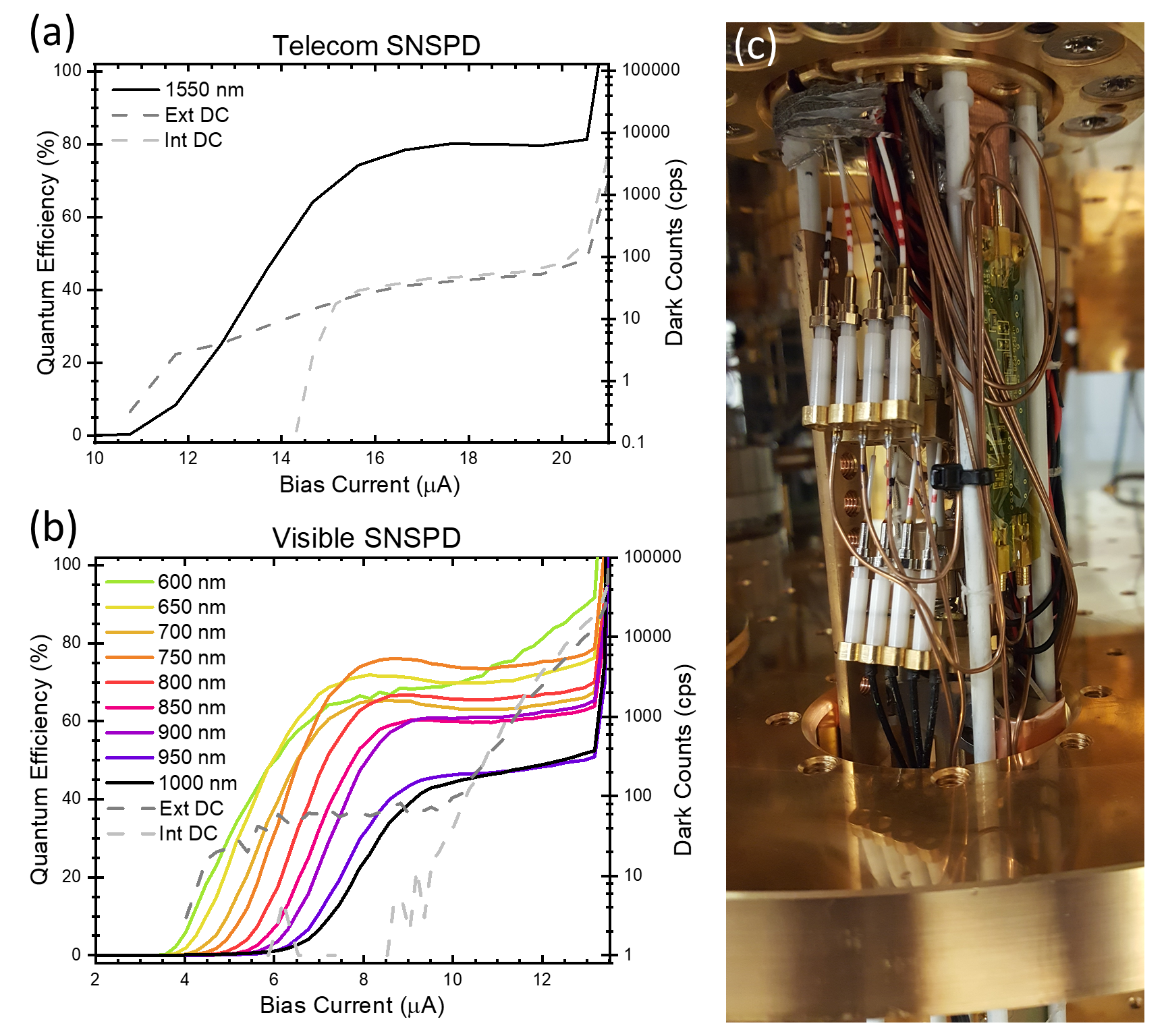}
    \caption{(a) Quantum efficiency (QE) at 1550 nm and dark counts (Int DC = intrinsic dark counts with fiber capped, Ext DC = extrinsic dark counts with fiber run across the lab) versus bias current for a telecom (1300-1600 nm) Quantum Opus, LLC detector mounted on the still-plate of the warm-insertable probe with the QE calibrated based on factory measurements. (b) QEs for a range of wavelengths (600 - 1000 nm) and dark counts versus bias current for a visible SNSPD mounted on the still from Quantum Opus, LLC.  The QEs were calculated by normalizing to the known QE of an avalanche photodiode. (c) Photograph of the four telecom (red marked fiber) and four visible (black marked fiber) SNSPDs mounted on the still, half of them are connected to cryoamplifiers mounted to the 3 K plate above them.}
    \label{fig:SNSPD}
\end{figure}

Because our confocal microscope is designed to operate at temperatures of 10 mK-1 K, the integration of SNSPDs for single photon detection is more straightforward than in warmer optical microscopes. However, where integration of SNSPDs at mK temperatures in the microscope itself could be plausible for a fiber-coupled microscope, here SNSPDs were installed on a separate warm-insertable 50.2 mm diameter probe. Eight Quantum Opus SNSPDs in total are installed and heatsunk to the still at 800 mK. Four are optimized for broadband visible single photon detection.  These visible SNSPDs utilize a 30 micron active area, and they are coupled to 30 micron graded index multimode fiber that is accessible through hermetic fiber connections mounted to the 300 K plate.  The other four SNSPDs are optimized for broadband telecom operation, and they are accessible through SMF28 single mode fiber at the 300 K plate. Then, as illustrated in Fig.~\ref{fig:schematic}, a flip mirror on the breadboard is used to route photons collected from the free-space optical microscope back to the SNSPDs.  Four of the SNSPDs (two visible and two telecom) utilize cryoamplifiers mounted to the 3 K plate of the probe in order to provide improved jitter and a path toward limited photon number resolution~\cite{cahall2017multi}. While multimode SNSPDs offer improved collection efficiency for quantum optical microscopies where the collected light is frequently not well described by a TEM$_{00}$ mode, they also exhibit oscillations in the readout pulse that are not typically observed in single mode devices~\cite{marvinney2020waveform}.  However, at least for 30 micron devices, the oscillations do not affect the performance of the SNSPDs in a detrimental way, and the phase of the oscillations can be used to coarsely infer the position that each photon was detected on the detector~\cite{marvinney2020waveform}. Figure~\ref{fig:SNSPD} illustrates the detector installation on our warm-insertable probe and some basic performance parameters for the visible and telecom SNSPDs. Notably, the maximum fields at the sample of 6 T in z and 1 T in x and y result in fields at the SNSPDs of 125 G  and 55 G respectively. While the bias current at which the SNSPDs begin to turn on and at which the dark counts start to run away is slightly dependent on magnetic field at this scale, the device performance for bias currents at which the quantum efficiency is plateaued is independent of magnetic field. 

\section{Raman and photoluminescence spectroscopies}

Polarization and field-resolved Raman and photoluminescence spectroscopies relying on the optics train illustrated in Fig.~\ref{fig:schematic} are ongoing. These measurements rely on either 532 nm off-resonant excitation or tunable resonant excitation provided by the C-wave optical parametric oscillator. Polarization-resolved field-dependent Raman and photoluminescence microscopies rely on the same polarization calibration described above.  The dichroic mirror illustrated in Figure 3 is mounted on a removable magnetic mount so that experiments can be quickly modified to incorporate (a) resonant excitation through a 90/10 beamsplitter, (b) 532 nm or higher energy excitation through a Semrock dichroic filter, or (c) 532 nm excitation through an Optigrate dichroic filter that enables low energy Raman and a combination of Stokes and anti-Stokes spectroscopies.

\begin{figure}[h]
\centering
    \includegraphics[width=.9\columnwidth]{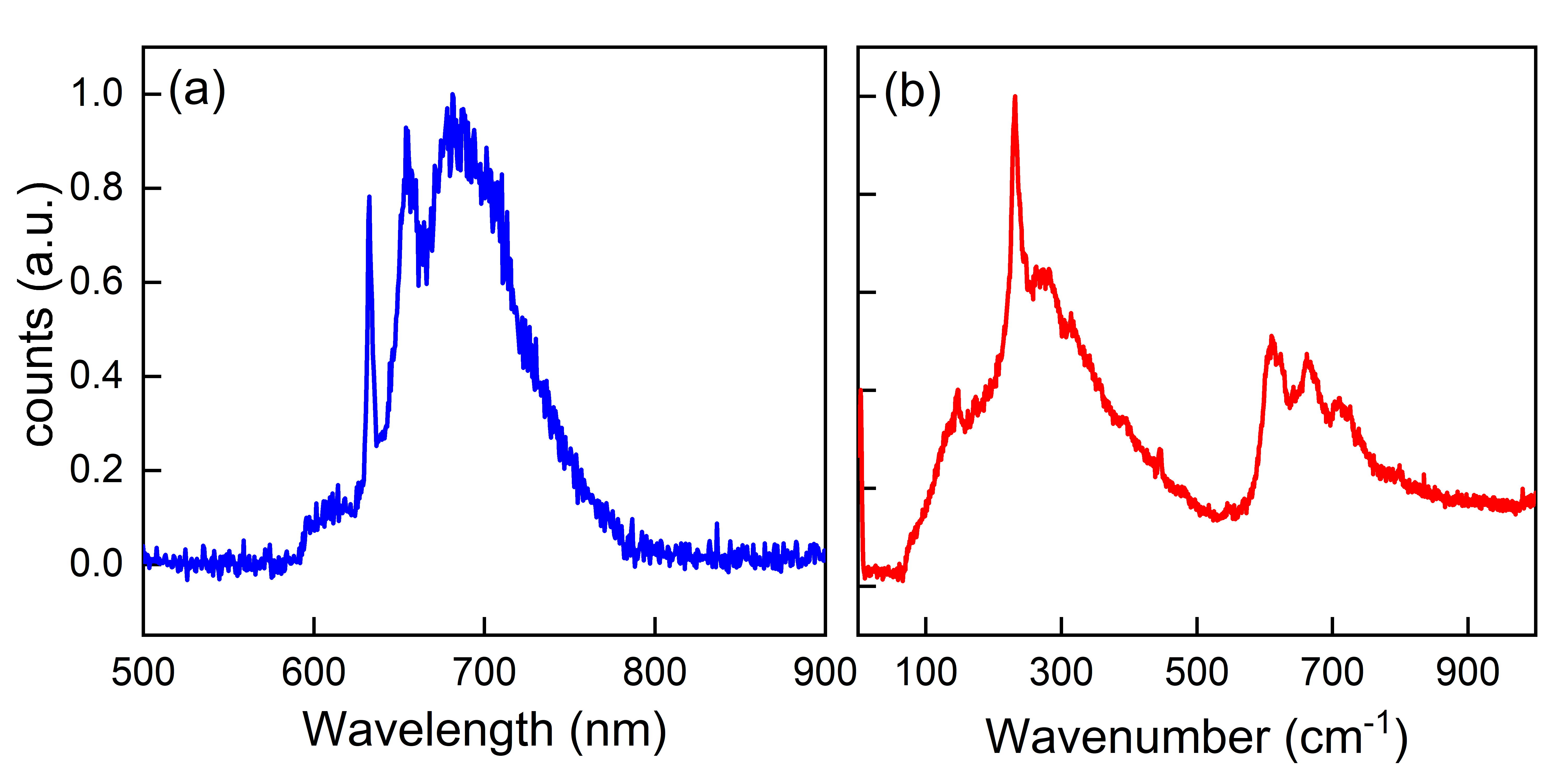}
    \caption{(a) Photoluminescence spectrum of ensemble NV centers in a type Ib nanodiamond sample acquired for 500 s with a 55 $\mu W$, 532 nm pump laser at a temperature of 55 mK.  Raman spectrum of a strontium titanate sample acquired for 3 hrs with a 1 mW, 532 nm pump laser at a temperature of 800 mK.}
    \label{fig:RamanPL}
\end{figure}

An Andor Kymera 193i spectrometer is used for basic Raman and photoluminescence spectroscopies after fiber collection on the breadboard.  Gratings of 150-2400 lines/mm provide spectroscopy with flexible spectral resolution and bandwidth.  However, compact spectrometers are inherently limited in their spectral resolution.  At 2400 lines/mm, spectral resolution of 0.1 nm is available within a small spectral band.  Spectroscopic characterization of high quality factor transitions therefore relies on two alternate approaches.  First, a fiber coupled spectrometer that combines a virtually imaged phase array etalon with a conventional diffraction grating (LightMachinery Hyperfine spectrometer) provides 3 pm spectral resolution across much of the visible spectrum.  Second, photoluminescence excitation (PLE) spectroscopies that leverage the 40 GHz mode-hop free tuning range and <1 MHz linewidth of the C-Wave enable ultrahigh resolution resonant excitation spectroscopies, with readout in the phonon sidebands of specific excitonic or defect transitions.  Combining mK photoluminescence spectroscopies (utilizing the Andor and LightMachinery spectrometers) with PLE spectroscopies is critical to a complete understanding of the excitation and relaxation processes in excitonic and defect-based light emitting systems. Figure~\ref{fig:RamanPL} illustrates prototypical Raman and photoluminescence spectra measured in this system.

As described above, the base temperature of these measurements is largely limited by the required laser power for a given measurement.  Thus, photoluminescence spectroscopies of single photon emitters rely on optimized collection efficiency in order to minimize the required laser power.  Resonant excitation generally yields substantially improved excitation efficiencies, resulting in lower laser power and reduced laser heating.  Raman spectroscopies of some layered 2D materials are particularly challenging as we approach the monolayer limit.  As the Raman signal drops below the photon shot noise limit, it becomes challenging to recover a measurable signal without increasing the sample temperature excessively.  Microscopies that rely on squeezed optical readout fields to push the readout noise floor below the photon shot noise limit~\cite{lawrie2019quantum,lawrie2020squeezing,de2020quantum,pooser2020truncated} could help to reduce the required laser power for optical microscopies that are otherwise constrained by the photon shot noise limit, and truncated nonlinear interferometries~\cite{lawrie2020squeezing,pooser2020truncated} could help to reduce the optical power delivered to the sample by additional orders of magnitude, potentially enabling Raman spectroscopy near the base temperature of the dilution refrigerator.  However, squeezing is generally very sensitive to optical loss.  Thus any microscope that relies on a squeezed readout field must be well optimized to minimize loss and optimize the available quantum advantage.

\section{Relaxometry}
The spin dynamics of defects coupled to different quantum systems serve as an essential probe of fundamental excitations in quantum materials\cite{candido2020predicted,du2017control}.  Spin relaxometry and spin coherence measurements require a combination of optical, microwave, and readout pulses that must be optimized to realize useful SNRs without inducing excess heating. Defects with triplet ground states and a corresponding intersystem crossing relaxation pathways only require an off-resonant 532 nm excitation for spin state initialization and readout~\cite{Barry2020}. On the other hand, defects with spin-orbit selection rules require excitation pulses resonant with the spin state to be initialized or readout~\cite{Sukachev2017,Trusheim2020}. An additional recharging pulse is often used prior to resonant excitation pulses as well to assure the defect is repumped into the correct charge state. Thus, two color pulse sequences that combine the off-resonant 532 nm excitation and a resonant OPO pulse can be designed and controlled by a Swabian pulsestreamer that modulates the two AOMs illustrated in Fig.~\ref{fig:schematic}. The pulse streamer is also used to control the output of a Keysight M8195A arbitrary waveform generator, and a Swabian Timetagger Ultra 8 is used to timetag all photons detected on the integrated SNSPDs or APDs.

\section{Conclusion}

Optical microscopies and spectroscopies are increasingly critical to the characterization of mesoscale properties of quantum materials and quantum devices at milliKelvin temperatures.  Fiber inserts have proven to be a useful platform for such optical spectroscopies when free-space access is infeasible~\cite{sladkov2011polarization,macdonald2015optical}, but free-space optical microscopes generally provide substantially greater flexibility in terms of imaging and polarization control than fiber-coupled microscopes. As described above, combining stick slip positioners for coarse sample positioning with galvonometer controlled steering from outside the dilution refrigerator is essential to achieving high spatial resolution scanning at low temperatures without excess heating or rf noise.  Substantial calibration is needed for linearly polarized experiments performed in large magnetic fields parallel to the optical axis, but such calibration is straightforward.  By contrast, circular dichroism experiments can be performed with relatively minimal calibration in large magnetic fields with appropriate experimental design.  Some tradeoff is always necessary between vibration levels, sample base temperature, cooling power, optical access, and ease of sample transfer, but we believe the design we have described here provides a reasonable compromise within this parameter space.  Ongoing experiments are targeting scanning laser and scanning single photon microscopies of superconducting devices, photoluminescence and Raman microscopies of transition metal dichalcogenides and candidate quantum spin liquids, and characterization of the spin dynamics of hybrid quantum systems.

\section{Acknowledgments}
This research was sponsored by the U. S. Department of Energy, Office of Science, Basic Energy Sciences, Materials Sciences and Engineering Division. The installation of the dilution refrigerator and the initial optical design was supported by the Laboratory-Directed Research and Development Program of Oak Ridge National Laboratory, managed by UT-Battelle, LLC for the U.S. Department of Energy.  M.A.F. gratefully acknowledges student research support by the Department of Defense (DoD) through the National Defense Science \& Engineering Graduate Fellowship (NDSEG) and NSF award DMR-1747426. C.E.M gratefully acknowledges postdoctoral research support from the Intelligence Community Postdoctoral Research Fellowship Program at the Oak Ridge National Laboratory, administered by Oak Ridge Institute for Science and Education through an interagency agreement between the U.S. Department of Energy and the Office of the Director of National Intelligence. Additional thanks go to Leiden Cryogenics, Nanomagnetics Instruments, and Quantum Opus LLC. for their support in the development of this system.


%

\end{document}